\documentclass[aps,prl,letterpaper,amsmath,amssymb,superscriptaddress,nofootinbib,twocolumn]{revtex4-2}
\usepackage{graphicx}
\usepackage[usenames]{color}
\usepackage[colorlinks=true,linkcolor=blue,citecolor=blue,anchorcolor=green,urlcolor=blue]{hyperref}
\usepackage{subfigure}
\usepackage{color}
\usepackage{mathrsfs}
\def\be{\begin{equation}}
\def\ee{\end{equation}}
\def\ba{\begin{aligned}}
\def\ea{\end{aligned}}
\usepackage[normalem]{ulem}

\usepackage{booktabs}
\usepackage[percent]{overpic}
\usepackage{tabularx}
\usepackage{multirow}
\usepackage{CJK}
\usepackage{booktabs}
\usepackage{makecell}

\begin{document}
\title{Spin dynamics and dark particle in a weak-coupled quantum Ising ladder with $\mathcal{D}_8^{(1)}$ spectrum}

\newtheorem{theorem}{Theorem}[section]
\newtheorem{lemma}[theorem]{Lemma}
\author{Yunjing Gao}
\affiliation{Tsung-Dao Lee Institute,
Shanghai Jiao Tong University, Shanghai, 201210, China}

\author{Xiao Wang}
\affiliation{Tsung-Dao Lee Institute,
Shanghai Jiao Tong University, Shanghai, 201210, China}

\author{Ning Xi}
\affiliation{Department of Physics and Beijing Key Laboratory of Opto-electronic
Functional Materials and Micro-nano Devices, Renmin University of China,
Beijing 100872, China}
\affiliation{CAS Key Laboratory of Theoretical Physics, Institute of Theoretical Physics, Chinese Academy of Sciences, Beijing 100190, China}

\author{Yunfeng Jiang}
\affiliation{School of Phyiscs and Shing-Tung Yau Center, Southeast University, Nanjing 210096, China}

\author{Rong Yu}
\altaffiliation{rong.yu@ruc.edu.cn}
\affiliation{Department of Physics and Beijing Key Laboratory of Opto-electronic
Functional Materials and Micro-nano Devices, Renmin University of China,
Beijing 100872, China}
\affiliation{Tsung-Dao Lee Institute,
Shanghai Jiao Tong University, Shanghai, 201210, China}
\affiliation{Key Laboratory of Quantum State Construction and Manipulation (Ministry of Education), Renmin University of China, Beijing, 100872, China}

\author{Jianda Wu}
\altaffiliation{wujd@sjtu.edu.cn}
\affiliation{Tsung-Dao Lee Institute, Shanghai Jiao Tong University, Shanghai, 201210, China}
\affiliation{School of Physics \& Astronomy, Shanghai Jiao Tong University, Shanghai, 200240, China}
\affiliation{Shanghai Branch, Hefei National Laboratory, Shanghai 201315, China}
\date{\today}

\begin{abstract}
Emergent Ising\(_h^2\) integrability is anticipated in a quantum Ising ladder
composed of two weakly-coupled
critical transverse field Ising chains.
The system is remarkable for including eight types of massive relativistic particles,
with their scattering matrix and mass spectrum characterized by the $\mathcal{D}_8^{(1)}$ Lie algebra.
In this article,
by computing the spin dynamical structure factors following analytical form factor approach, 
we clearly identify dispersive single-particle excitations of (anti-) soliton 
and breathers as well as their multi-particle continua in the spectra,
which is further confirmed by the numerical simulations.
We show that the selection rule inherent in the parity and
topological charge of the theory,
causes a significant result that charge-parity-odd particles,
termed as dark particles,
cannot be directly excited from the ground state through any local or quasi-local operations.
This in turn 
suggests the long lifetime of the lightest dark particle.
\end{abstract}
\maketitle

\paragraph*{Introduction.---}
Emergent conformal invariance and integrability manifest
in a variety of critical two-dimensional (2D) classical statistical
models \cite{Polyakov1970,BELAVIN1984333} and 1D quantum critical systems \cite{sachdev_2011}.
Building upon these foundations,
studies on integrable deformation and
higher-dimensional systems gained widespread attention \cite{SOS,coupleCFT,ladderBA1}.
When described by an integrable field theory,
the accompanying algebraic structure \cite{byb,ATFT} provides a guiding framework for studying the particle excitations as well as spectral characteristics of the corresponding lattice model.
A family of paradigmatic models originates in the transverse field Ising chain (TFIC) \cite{Isingphase,PhysRevLett.120.207205,PhysRevB.97.245127,Yang_2023},
where
fruitful quantum critical physics
and elegant quantum integrability have been revealed.
Conformal field theory (CFT) with central charge $1/2$ emerges when the TFIC
is tuned to its quantum critical point (QCP) \cite{BELAVIN1984333,IsingCFT}.
The perturbation of longitudinal field along Ising spin direction
further drives the TFIC into the quantum $E_8$ integrable model \cite{ZamE8},
in which the dynamical spectrum of the system is
controlled by the $E_8$ exceptional Lie algebra.
Experimentally, the $E_8$ physics was first proposed in the quasi-1D magnetic material 
CoNb$_2$O$_6$ \cite{Coldea} and has been recently confirmed in another quasi-1D antiferromagnet
BaCo$_2$V$_2$O$_8$ inside its 3D magnetic 
ordered phase upon transverse-field tuning \cite{E8,PhysRevB.101.220411,dispersion}.

Another set of integrable systems has been discovered within a category of coupled minimal CFTs \cite{coupleCFT}.
The quantum Ising ladder \cite{Ramos_PRB_2020} formed by two weakly coupled
quantum critical TFICs is effectively
described by the Ising$_h^2$ integrable field theory (IIFT) containing eight types of particles,
whose scattering matrix and mass spectrum are organized
by the $\mathcal{D}_8^{(1)}$ Lie algebra \cite{coupleCFT}.
However, whether this predicted $\mathcal{D}_8^{(1)}$ spectrum can be observed 
in the quantum Ising ladder is still open 
given that the spin dynamical structure factors (DSFs) of the model have not been generally discussed.
This motivates us to study the spin dynamics of
the quantum Ising ladder model.

In this article,
first we briefly review how the IIFT emerges from the quantum Ising ladder.
By identifying selection rules originated from global properties of the IIFT,
we show the existence of ``dark particles'' in the system,
which are charge-parity ($\mathcal{C}$)-odd,
and cannot be directly excited from the ground state 
through any local or quasi-local operations.
In particular,
the lightest one is forbidden from spontaneous decay as being shielded by $\mathcal{C}$ and the gap.
Moreover,
spin DSFs with zero and finite transfer momentum
are determined via an analytical form factor approach and numerical calculations.
Relativistic particle dispersions are confirmed 
with different particle channels distinguished clearly,
where exotic single (anti-)soliton excitation
is  clearly identified.
Numerical zero-temperature DSF results show clearly 
the absence of spectral weights for all local spin components 
at excitation energies corresponding to the predicted ``dark particles''.
The dark property is preserved even when deviating from the integrable point 
and can possibly be realized in magnetic materials, Rydberg arrays and etc.

\paragraph*{Model and bosonization.---}

\label{sec:Bosonization}
Consider a quantum Ising ladder composed by
two weakly coupled quantum critical TFICs

\begin{figure}[tp]
    \centering
    \includegraphics[width=0.36\textwidth]{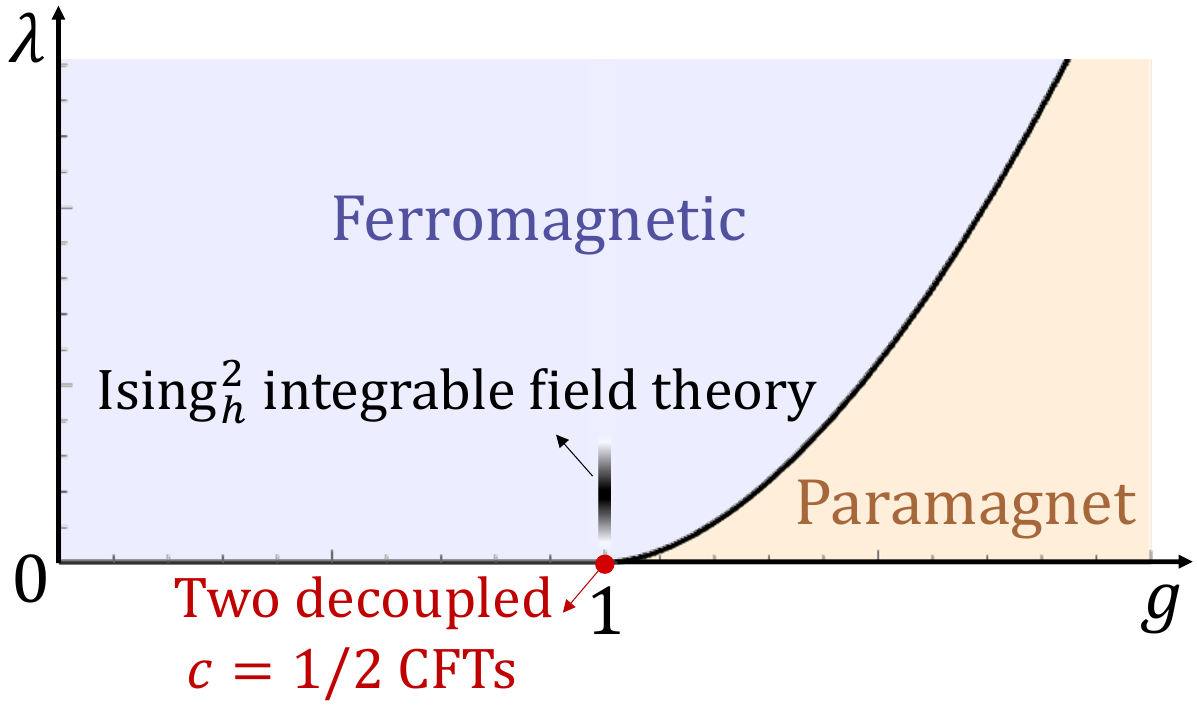}
    \caption{Illustration of phase diagram for the transverse-field Ising ladder. $\lambda$ and $g$ denote the interchain coupling and the transverse magnetic field, respectively. $g=1$ corrresponds to the critical point of two decoupled Ising chains, which can be described by two $c=1/2$ CFTs. The Ising$^2_h$ integrability emerges when a small $\lambda$ is turned on.}
    \label{fig:ill}
\end{figure}

\begin{subequations}
\begin{align}
&H=H_{\text{TFIC}}^{(1)}+H_{\text{TFIC}}^{(2)}+\lambda\sum_{i=1}^N \sigma^{z(1)}_i\sigma^{z(2)}_i,
\label{eq:ladderH}
\\
&H_{\text{TFIC}}^{(1,2)}=-J\left(\sum_{i=1}^{N-1}\sigma_i^{z(1,2)}\sigma_{i+1}^{z(1,2)}+g\sum_{i=1}^{N}\sigma_i^{x(1,2)}\right),
\label{eq:cTFIC}
\end{align}
\end{subequations}
with Pauli matrix at site $j$, $\sigma_j^{\mu(1,2)}=2S_j^{\mu(1,2)}\, (\mu=x,y,z)$,
intrachain coupling $J$, and interchain coupling $\lambda$.
Here $g=1$, quantum critical point for each chain.
Then in the scaling limit, i.e., lattice spacing 
$a\to0$, $\lambda \to 0$ but with finite $\lambda/a$,
the IIFT emerges from the two weakly-coupled TFICs [Eq.~\eqref{eq:ladderH}]
 (Sec.~A of the Supplemental Material (SM) \cite{SuppMat}).
In the scaling limit
the local Ising order operators $\sigma^{z(1,2)}_j$ become Ising field $\sigma^{(1,2)}(x)$, and $\sigma^{x(1,2)}_j$ are recasted as
energy density operators $\epsilon^{(1,2)}(x)$ \cite{KADANOFF}.
For later discussion
disorder operator $\mu^{(1,2)}(x)$ is introduced 
as the scaling limit of $\mu_j^{(1,2)}=\prod_{k=0}^{j-1}\sigma^{x(1,2)}_k\, (x=ja)$.

In the decoupled case [Fig.~\ref{fig:ill}]
(also known as a special case of the Ashkin-Teller model \cite{Ashkinteller,SPT}),
each critical TFIC can be described by the central charge $1/2$ CFT 
with respect to free massless Majorana spinor $\psi^{(1,2)}=(\psi_R^{(1,2)},\psi_L^{(1,2)})^T$ \cite{Majorana}.
Given two copies of critical TFICs,
the two sets of Majorana spinors
can be combined into a Dirac spinior
$\chi = (\psi^{(1)}+i\psi^{(2)})/\sqrt{2}$
which can be further bosonized.
Following the bosonization rule $\chi_{L}\sim\exp(-i\phi_{L})$ and
$\chi_{R}\sim\exp(i\phi_{R})$,
the two decoupled TFICs can be mapped to a free boson field theory 
with the free bosonic field $\phi(x,t)=\phi_R(x-t)+\phi_L(x+t)$ \cite{2isingboson,banks}
(see also Sec.~A of the SM \cite{SuppMat}).
The operator correspondences follow \cite{KADANOFF,2isingboson},
\be\ba 
&\sigma^{(1)}\sigma^{(2)}\sim\cos\frac{\phi}{2},\quad 
\mu^{(1)}\mu^{(2)}\sim\sin\frac{\phi}{2},\\
&\epsilon^{(1)}+\epsilon^{(2)}\sim\cos\phi,\quad \epsilon^{(1)}-\epsilon^{(2)}\sim\cos\Theta,
\label{eq:bosonization}
\ea\ee
where  $\Theta(x,t)=\phi_R(x-t)-\phi_L(x+t)$ is the dual field of $\phi$.

When the interchain coupling $\lambda$ turns on,
Eq.~\eqref{eq:bosonization} implies that
the interchain spin interaction introduces the $\cos(\phi/2)$ term into the free model.
As such in the scaling limit the lattice model Eq.~\eqref{eq:ladderH} converts to
the IIFT,
\be
\mathcal{A}=\int dxdt\left\{\frac{1}{16\pi}(\partial\phi)^2+\Lambda\cos\frac{\phi}{2}\right\},
\label{eq:SGaction}
\ee
with rescaled interchain coupling constant $\Lambda$. It is worth to mention that
Eq.~\eqref{eq:SGaction} appears the same as a reflectionless sine-Gordon model (SG$_{1/8}$) \cite{Coleman}.
The two models are distinctive in that the former is defined
on a $\mathbb{Z}_2$ orbifold while the latter is compactified on a circle \cite{coupleCFT}.
Nonetheless, both IIFT and SG$_{1/8}$ are quantum integrable systems
that possess excitations
associated with the $\mathcal{D}_8^{(1)}$ algebra \cite{massalgebra},
with totally 8 types of particles
including 6 breathers $B_n$ with masses $m_{B_n}=m_{B_1} \sin(n\pi/14)/\sin(\pi/14),(n=1,\dots,6)$,
one soliton ($A_{+1}$) and one antisoliton ($A_{-1}$)
with mass $m_{A_{+1}}= m_{A_{-1} }= m_{B_1}/(2\sin(\pi/14))$.

\paragraph{Form factors and selection rules.---}
%\paragraph{S-matrices, form factors and selection rules.---}
%\paragraph*{Form factors of the Ising$_{h}^2$ model.---}
In the context of a (1+1)D integrable quantum field theory (IQFT),
the form factor (FF) $
F_{\hat{\mathcal{O}}}^{P_1,\dots,P_n}(\theta_1>\theta_2>\dots>\theta_n)=\langle 0|\hat{\mathcal{O}}|P_1(\theta_1)P_2(\theta_2)
\dots P_n(\theta_n)\rangle$ of a local observable $\hat{\mathcal{O}}$ can in principle
be derived from the form factor
bootstrap scheme \cite{KAROWSKI1978455, Smirnov},
with the $j$-th particle of type $P_j$ carrying rapidity $\theta_j$.
Here the translational invariant asymptotic in-state (the ket) is an eigenstate 
of the corresponding integrable
Hamiltonian with energy $E=\sum_{j=1}^n m_{P_j}c^2 \cosh\theta_j$ and momentum
$q=\sum_{j=1}^n m_{P_{j}}c \sinh\theta_j$ \cite{Mussardobook},
where the speed of ``light''  $c=1$ in the field theory framework.

The FFs play a central role in studying spin DSF.
The DSF for a local observable
$\hat{\mathcal{O}}$ with transfer momentum $q$ and transfer energy $\omega$
$(\hbar=1)$ is given by
$D_{\hat{\mathcal{O}}}(q,\omega)=\int dxdt\langle\hat{\mathcal{O}}(x,t)
\hat{\mathcal{O}}(0,0)\rangle e^{i\omega t}e^{-i q x}$.
In the context of IQFT, it follows
\be
\ba
&D_{\hat{\mathcal{O}}}(q,\omega)=\sum_{n=1}^{\infty}\sum_{\{P_1 \cdots P_n\}}\int\frac{\mathscr{D} \theta}{\mathscr{A} (2\pi)^n}\left|F_{\hat{\mathcal{O}}}^{P_1\cdots P_n}(\theta_1,\cdots,\theta_n)\right|^2\\
&\times\delta\left(\omega-\sum_{l=1}^nm_{P_l}\cosh\theta_l\right)\delta\left(q-\sum_{l=1}^nm_{P_l}\sinh\theta_l\right),
\label{DSF}
\ea
\ee
where $\mathscr{D}\theta = \prod_{i=1}^{n}d\theta_i$ and
$\mathscr{A}=\prod_{l\in\{P_i\}} n_{l}!$ with $n_l$ satisfying $\sum_{j}n_j=n$,
which counts particle number of type $l$ in configuration $\{P_i\}$.
Explicitly,
for single particle channels,
\be
D_{\hat{\mathcal{O}}}^{P_1}(\omega,q)=\frac{2\pi}{\sqrt{m_{P_1}^2+q^2}}
|F^{P_1}_{\hat{\mathcal{O}}}|^2\delta\left(\omega-\sqrt{m_{P_1}^2+q^2}\right).
\label{single}
\ee
And for two-particle channels,
\be
D_{\hat{\mathcal{O}}}^{P_1,P_2}(\omega,q)=\frac{1}{1+\delta_{P1, P2}}\frac{|F_{\hat{\mathcal{O}}}^{P_1,P_2}
(\theta_1,\theta_2)|^2}{m_{P_1}m_{P_2}\sinh(\theta_1-\theta_2)}
\ee
with $e^{\theta_1}=[\sqrt{(m_{P_2}^2-m_{P_1}^2-\omega^2+q^2)^2-4m_{P_1}^2(\omega^2-q^2)}-(m_{P_2}^2-m_{P_1}^2-
\omega^2+q^2)]/[2m_{P_1}(\omega-q)]$ and $e^{\theta_{2}}=(\omega+q-m_{P_1} e^{\theta_1})/m_{P_2}$.

Spin excitations captured by non-vanishing DSF channels originates from
global properties of the IIFT.
Considering transverse spin dynamics,
we focus on the FFs of $\cos{\phi}$
and $\cos{\Theta}$ ($\exp(\pm i\phi)$, $\exp(\pm i\Theta)$).
In terms of the SG model terminologies,
the breathers carry zero topological charge ($Q$),
while the soliton and the antisoliton carry $Q = +1$ and $-1$, respectively.
Moreover,
$Q$ remains conserved under $e^{\pm i\phi}$
but shifts by $\pm 1$ through $e^{\pm i \Theta}$ \cite{chargeraising},
which results in the first selection rule based on the total $Q$ of the asymptotic state.
$e^{\pm i\phi}$ connects vacuum with an asymptotic state of $Q=0$,
where soliton and anti-soliton must appear in pairs.
Conversely,
state combined with $e^{\pm i\Theta}$ contains odd number 
of soliton(s) and anti-soliton(s) in total with net $Q = \pm 1$.

The second selection rule originates from
the charge conjugation (parity) operation $\mathcal{C}$ 
that acts as $\mathcal{C}\phi\,\mathcal{C}^{-1}=-\phi$ \cite{bookBosonization, Schroer}.
Correspondingly,
$\mathcal{C}|A_{\pm 1}(\theta)\rangle=|A_{\mp 1}(\theta)\rangle$ and
$\mathcal{C}|B_n(\theta)\rangle=(-1)^n|B_n(\theta)\rangle\,(n=1,2,\cdots,6)$.
The vacuum state $|0\rangle$ is $\mathcal{C}$-invariant.
Henceforce,
we omit unnecessary $\theta$ without causing confusion.
As $Q\neq 0$ states  do not possess well-defined $\mathcal{C}$-parity,
they are free from $\mathcal{C}$ selection rules.
Consequently,
non-vanishing form factors include:
$\langle 0|\cos\phi|B_{n_1}\cdots B_{n_N}A_{s_1}\cdots A_{s_M}\rangle$ ($M,N\in\mathbb{N}$)
with $\sum_{i=1}^N n_i$ even, $\sum_{i=1}^Ms_i=0$;
$\langle0|\exp(\pm i\Theta)|B_{n_1}\cdots B_{n_{N}}A_{s_1}\cdots A_{s_{M}}\rangle$ with $\sum_{i=1}^Ms_i=\pm1$.

Furthermore,
scattering matrices in the Ising$_h^2$ and SG$_{1/8}$ models differ by a minus sign
in $A_{\pm 1}A_{\pm 1\;{\text or}\; \mp 1}$ scatterings \cite{coupleCFT}.
Such difference only causes minor modifications for form factors
in the SG$_{1/8}$
 \cite{ZamZam,Lukyanov1995,Lukyanov1997,chargeraising,Takacs2010,Takacs2011,Palmai2012}. 
For instance,
the FF for a pair of soliton-antisoliton
in the IIFT follows by,
\be\ba
&\langle 0|e^{i\phi}|A_{+1}(\theta_1)A_{-1}(\theta_1-\alpha)\rangle=\mathcal{E}\sinh\frac{\alpha}{2}\frac{4e^{7\alpha/2}g(\alpha)}{-\sinh(7\alpha)/7}\\
&\times\left(\cosh\frac{\alpha}{2}\cot^2\frac{\pi}{14}\cot\frac{\pi}{7}+\cosh\frac{3\alpha}{2}\cot\frac{\pi}{14}\cot\frac{\pi}{7}\cot\frac{3\pi}{14}\right).
\ea\ee
with $g(\alpha) =
i\sinh\left(\frac{\alpha}{2}\right)e^{\int_0^{\infty}\frac{dt}{t}\frac{\sinh^2[t(1-i\alpha/\pi)]
\sinh[t(\xi-1)]}{\sinh(2t)\cosh(t)\sinh(t\xi)}}$ and normalization constant $\mathcal{E}$.
General expressions of form factors of the Ising$_h^2$ model can be found in Sec.~B of the SM \cite{SuppMat}.

\begin{figure}[t]
    \centering
    \includegraphics[width=0.46\textwidth]{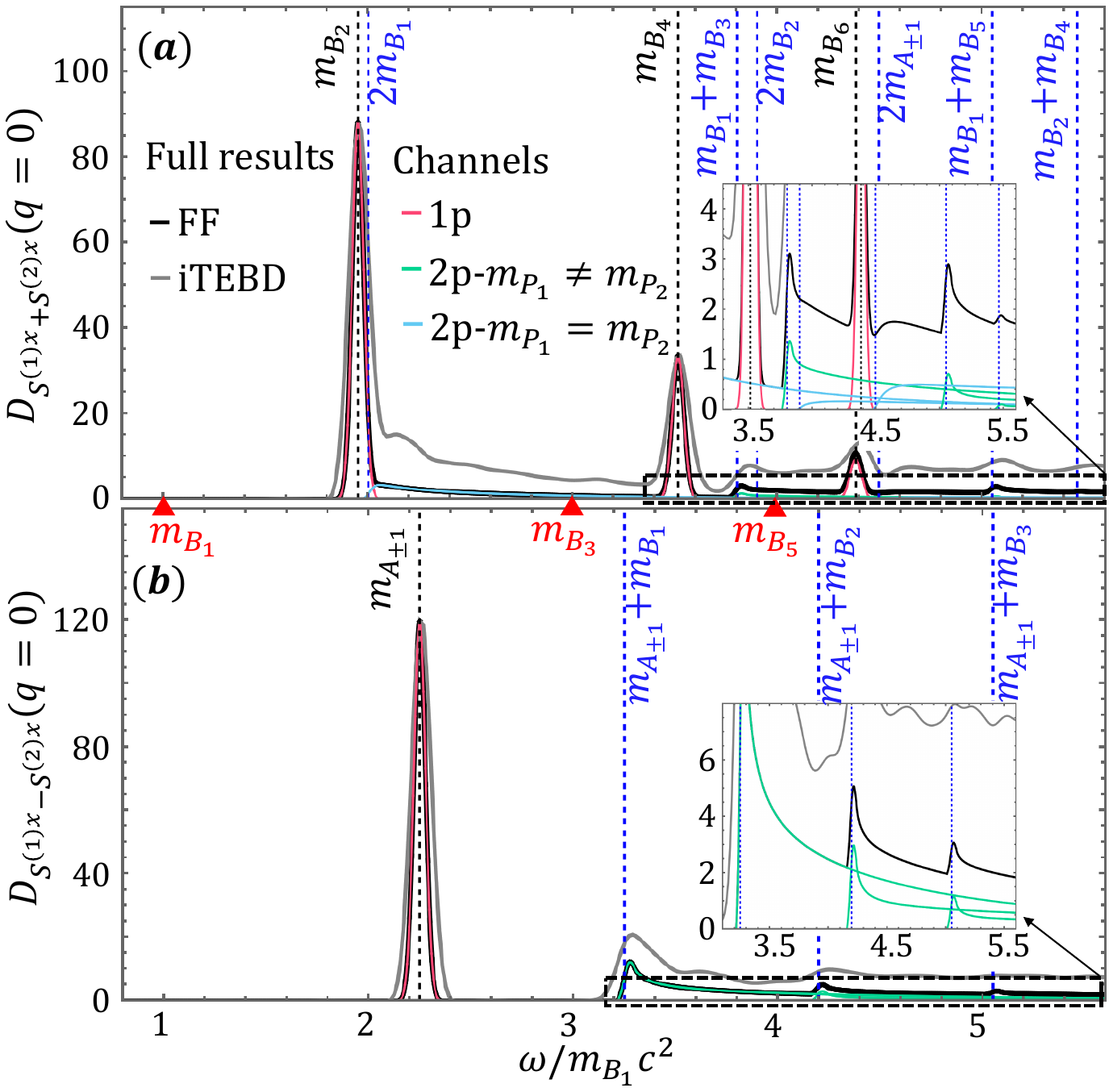}
    \caption{Spin DSFs $D_{S^{(1)x}\pm S^{(2)x}}(\omega)$ at lattice zone center $q=0$ determined
    from the analytical FF approach and the iTEBD simulation with $\lambda=0.1J$ (see \cite{SuppMat}).
    Contributions from different channels determined from the FF calculation are also present. 
    $1$p and $2$p are short notations for one-particle and two-particle.
    Vertical dashed lines illustrate the energy of
    single particle (rest masses) (black)
    and two particle thresholds (blue) with non-vanishing spectral weight,
    where the high energy regions are zoomed in.
    Red triangles mark energies of single $B_{1,3,5}$.
    The analytical results are normalized by the maximum values obtained from numerical calculations.
    }
    \label{Fig:cq0w}
\end{figure}

\paragraph*{Spin DSFs and dark particles.---}
\begin{figure*}[htp]
    \includegraphics[width=0.8\textwidth]{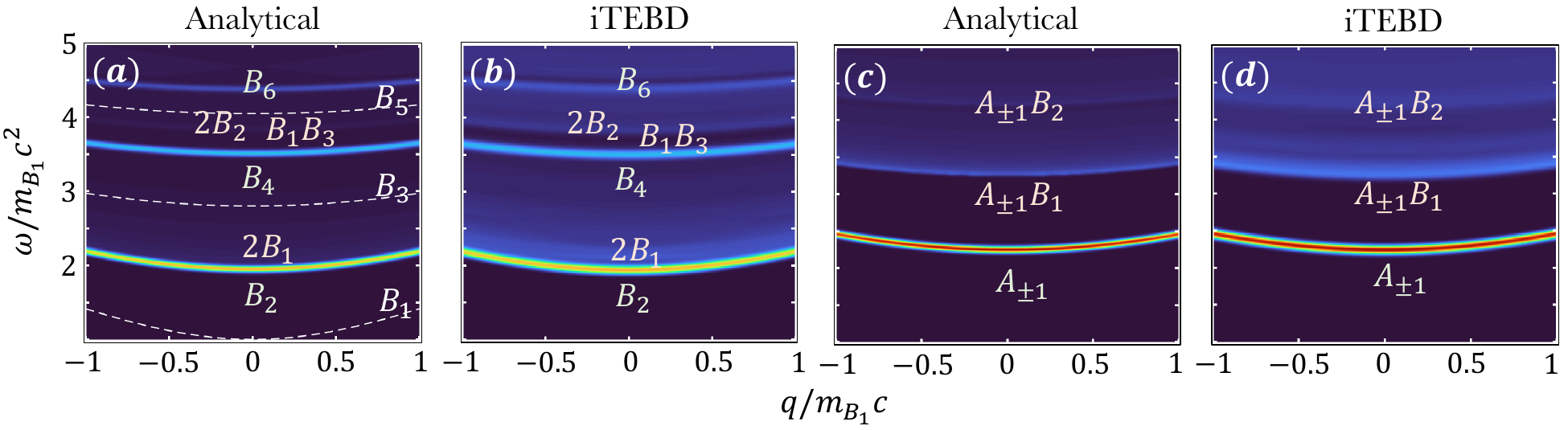}
    \caption{Spin DSFs $D_{S^{(1)x}+ S^{(2)x}}(\omega,q)$ in $(a,c)$ and $D_{S^{(1)x}-S^{(2)x}}(\omega,q)$ in $(b, d)$.
    $c=1$ in the analytical derivation and $c=1.5283[Ja/\hbar]$ for iTEBD calculation with $\lambda=0.1J$ determined by fitting the dispersion.
    The spectra weight for analytical results are normalized by aligning the maximum value with that obtained from numerical calculation for convenience. The white dashed lines in (a) 
     illustrate theoretical dispersions of $B_{1,3,5}$ for reference.}
    \label{Fig:finiteq}
\end{figure*}
With selection rules encoded in the FFs,
the DSF for $S^{(1)x}\pm S^{(2)x}$ channels are obtained.
Meanwhile we carry out the infinite time-evolving block decimation (iTEBD) calculations \cite{Vidal,SCHOLLWOCK2011} (Sec.~E of the SM \cite{SuppMat})
for the lattice model [Eq.~(\ref{eq:ladderH})] with $\lambda = 0.1J$.
All results are consistent with the selection rules.
The spectral weights for
single-$\mathcal{D}_8^{(1)}$ particles [Eq.~(\ref{single})] decay
fast with increasing particle's mass [Fig.~\ref{Fig:cq0w}].
Interesting features can also be found in the spectra continua, 
i.e., multi-particle channels.
For non-vanishing spectral weight in the two-particle channel,
we require $\omega\geq\omega_0=m_{P_1}+m_{P_2}$.
%$\omega(q)\geq\omega_0(q) \equiv \sqrt{(m_{P_1}+m_{P_2})^2+q^2}$.
At the threshold $\omega_0$,
$D_{\cos\phi}^{P_1,P_2}$ and $D_{\cos\Theta}^{P_1,P_2}$
diverge as $1/\sqrt{\omega-\omega_0}$ for $m_{P_1}\neq m_{P_2}$ [Fig.~\ref{Fig:cq0w}],
while the divergence is cancelled by the form factor when $m_{P_1}=m_{P_2}$.
The edge behaviors echo similar observations found in
other quantum integrable field theories \cite{PhysRevB.103.235117,E7}.

Furthermore,
the relativistic particle dispersion derived from analytical calculations
[Fig.~\ref{Fig:finiteq} (a, c)] are corroborated with numerical results [Fig.~\ref{Fig:finiteq} (b, d)]
from the lattice model [Eq.~\eqref{eq:ladderH}].
Now the two particle threshold is obtained as $\omega_0(q)=\sqrt{(m_{P_1}+m_{P_2})^2+q^2}$,
with similar divergent behavior as in $q=0$.
To make comparison we need to properly re-scale
the obtained analytical and numerical spectra independently \cite{PhysRevB.106.125149,dispersion}, namely,
we set infrared (IR) cutoffs for the momentum and
energy as $(q_{\text{IR}}, \omega_{\text{IR}})\sim (m_{B_1}c, m_{B_1}c^2)$.
The momentum cutoff is chosen to
cover about $10\%$ range of the Brillouin zone
for the lattice model,
where the numerical and analytical results match very well.
$c$ for numerical result is obtained via fitting
$\omega = \sqrt{m_{B_2}^2c^4+q^2c^2}$ through the lowest branch
identified as the $B_2$ channel in the spectrum [Fig.~\ref{Fig:finiteq}].
In the case of finite transfer momentum,
the two-particle channels with $m_{P_1}\neq m_{P_2}$ again contribute distinguishable boundaries
for the corresponding spectra continua.
Experimentally,
local measurement of transverse spin is captured by $D_{S_j^{(1,2)x}} = (D_{S_j^{(1)x}+S_j^{(2)x}}+D_{S_j^{(1)x}-S_j^{(2)x}})/4$,
which corresponds to the sum of spectral weights in Figs.~\ref{Fig:cq0w}, \ref{Fig:finiteq}.

\begin{figure}[t!]
    \centering
    \includegraphics[width=0.43\textwidth]{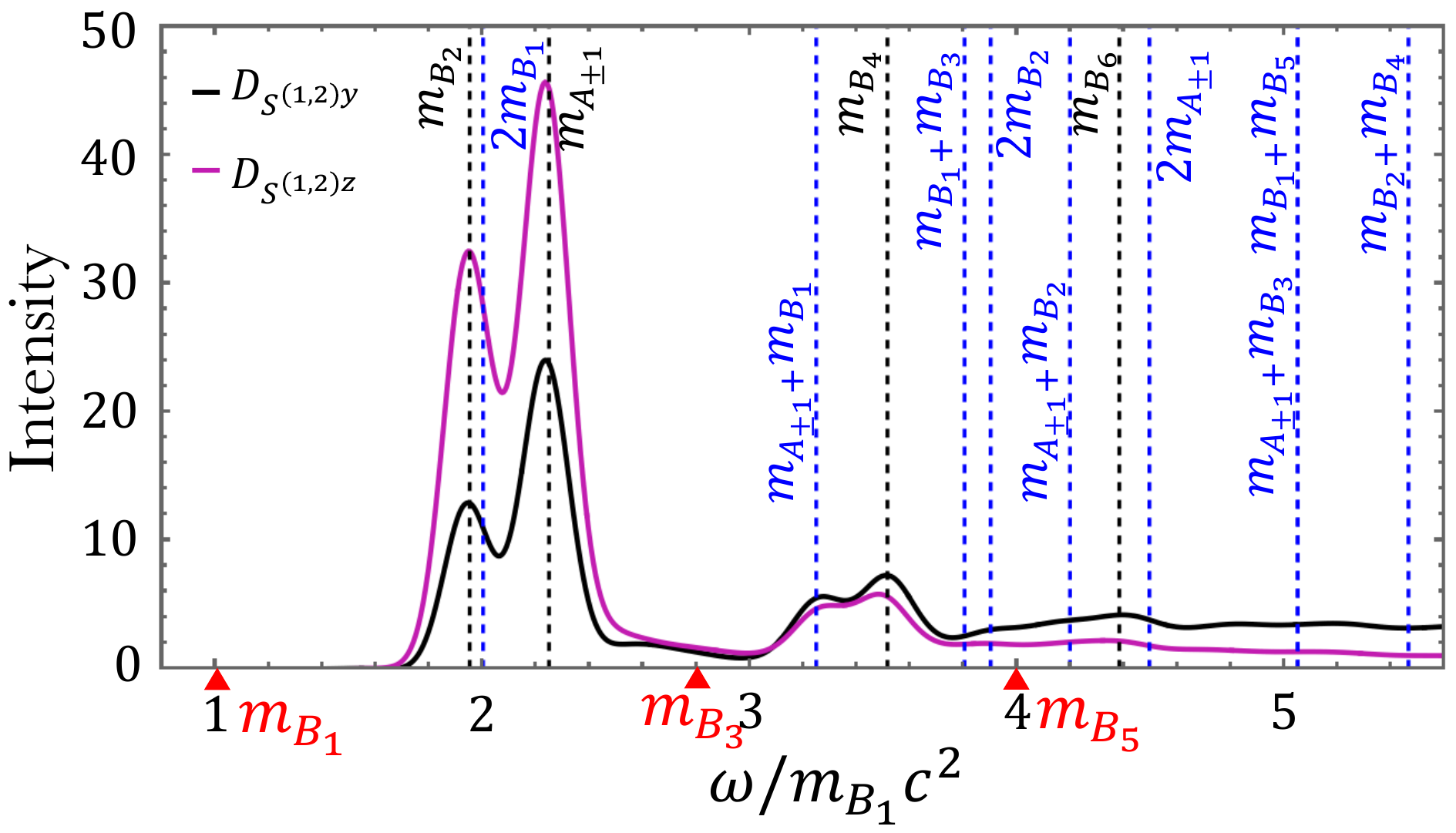}
    \caption{Spin DSFs of $S^{(1,2)y}$ and $S^{(1,2)z}$ at zone center calculated from iTEBD method (Sec.~E of SM \cite{SuppMat}) with $\lambda=0.1J$.
    $|B_{1,3,5}\rangle$ are absent. The black dashed lines denote single particle peaks.
    The blue dashed lines denote multi-particle energy thresholds
    echoing with those in Fig.~\ref{Fig:cq0w}.}
    \label{fig:yyzz}
\end{figure}

As discussed in the previous section,
transitions between single particle states $|B_{1,3,5}\rangle$ and $\mathcal{C}$-even
states (including the ground state) through $\sigma_j^{x}$ channel are forbidden.
The forbiddance is lifted for the disorder operator \cite{Schroer} $\mu^{(1)}(x)\mu^{(2)}(x)$ 
with the $\mathcal{C}$-parity odd.
This implies that to connect $\mathcal{C}$-odd neutral ($Q = 0$) states with the ground state,
global operation involving macroscopic number 
of $\sigma^x_j$'s from both chains is required.
Therefore,
we conclude that the lightest dark particle $|B_1\rangle$ ($\mathcal{C}$-odd)
is forbidden from spontaneous decay via vacuum fluctuations.
The parity-allowed decay channel $\mu^{(1)}(x)\mu^{(2)}(x)$
requires simultaneous operation on macroscopic number ($l$) of neighboring spins,
which, however, due to the gap, is exponentially cut off as $e^{-la/\xi_0}$ with vacuum fluctuation
being short spin
coherent length $\xi_0$ ($\sim\hbar c/m_1\approx 5a$ )
(typical parameters from 
magnetic materials \cite{toappear}).
Moreover,
Fig.~\ref{fig:yyzz} shows the absence of $B_{1,3,5}$ excitations 
through $S^{x,y,z}$ channels (Sec.~C of SM \cite{SuppMat}) by numerical simulations,
which agrees with the conclusion that global operation is required for these excitations.

\paragraph*{Conclusions and discussions.---}
Established on the emergent integrability for weak-interchain-coupled quantum critical Ising ladder,
we study the transverse component of the DSF spectrum by the form factor approach for the corresponding IIFT,
which are in a good agreement with numerical simulations.
The IIFT possesses 8 types of particles
with mass spectrum and scattering matrices
organized by the $\mathcal{D}_8^{(1)}$ algebra \cite{coupleCFT}.
The DSF shows clear single particle dispersions and two-particle continua.
Exotic single (anti)soliton excitation also exists in the spectrum,
which is beyond conventional understanding of pairwise domain wall excitations due to spin flip.
Selection rules originate in topological charge conservation and $\mathcal{C}$-symmetry are encoded in operators and the particles,
which are reflected by the spectrum.

It is worth discussing further the
existence of ``dark particle'' whose direct excitation from the ground state through local and quasi-local operations is forbidden.
Perturbation that turns the system away from the integrable point
will not change the dark nature of those magnetic excitations
as afore-discussed.
Typically,
let's consider the case that $g$ deviates from $1$.
The transverse field coupling terms can be bosonized into $\cos\phi$,
which is $\mathcal{C}$-even.
The dark property of $B_1$ is preserved as can be seen
from the perturbation scheme \cite{perturbation},
which leads to vanishing DSF spectra of perturbed $B_1$ channel due to the cancellation between soliton-antisoliton contribution and its charge conjugation (see Sec.~D of SM \cite{SuppMat}).
This argument also holds for perturbation coupled to $\sigma^{z}$.
These results are consistent with the iTEBD simulations shown in Fig.~S1,~S2 of SM \cite{SuppMat}.
Therefore,
the dark particle,
or more broadly, the low-lying dark magnetic modes remain inherently stable in an enlarged parameter region,
which is advantageous for practical application.

As for experimental observations of the $\mathcal{D}_8^{(1)}$ spectrum,
Ising ladders with weak rung interaction under transverse magnetic field should be an ideal platform.
As proposed in Ref.~[\onlinecite{toappear}],
several Ising-chain compounds may also serve as good candidates.
For instance,
though the pioneer work \cite{Coldea} on the quasi-1D magnetic material CoNb$_2$O$_6$
found evidence of $E_8$ spectrum with approximate golden ratio in energies of the two lowest excitations,
a recent THz measurement \cite{ZWang} revealed a number of additional excitation peaks beyond the $E_8$ integrable model.
As discussed in Ref.~[\onlinecite{toappear}],
tuning CoNb$_2$O$_6$ at a putative 1D QCP,
inside 3D order but near the 3D QCP \cite{Coldea,ZWang},
can possibly serve as a test bed for realizing the $\mathcal{D}_8^{(1)}$ 
spectrum.
A thermal activated detection on the dark particle has also been proposed \cite{thermal}.
Apart from quantum magnets,
$\mathcal{D}_8^{(1)}$ physics may also be explored 
within a finite system (Sec. F of SM \cite{SuppMat}), e.g., cold-atom simulations or by 
direct engineering in STM experiments.

\paragraph*{Acknowledgments.---}
We thank Linhao Li for helpful discussions at the early stage of this work.
This work is supported by 
the National Natural Science Foundation of China (Grant Nos.~12450004, 12274288, 12334008, 
12347138 and 12174441),
the Innovation
Program for Quantum Science and Technology Grant No. 2021ZD0301900,
the Natural Science Foundation of Shanghai with Grant No. 20ZR1428400,
and the National Key R\&D Program of China (Grant No. 2023YFA1406500).
We thank the High Performance Computing at RUC and ITP-CAS for the
technical support and generous allocation of CPU time.

\appendix
\newpage
\onecolumngrid
\setcounter{figure}{0}
\makeatletter
\renewcommand{\thefigure}{S\@arabic\c@figure}
\setcounter{equation}{0} \makeatletter
\renewcommand \theequation{S\@arabic\c@equation}
\section*{{Supplementary Material --- Spin dynamics and dark particle in a weak-coupled quantum Ising ladder with \(\mathcal{D}_8^{(1)}\) spectrum}}

\section{A. Bosonization of Ising ladder}
Consider a critical transverse field Ising chain (Eq.(1b) in the main text).
Following the Jordan-Wigner transformation,
spin operators can be mapped to fermionic operators $c_j^{\dagger}, c_j$ as \cite{Pfeuty}
\be
\sigma^z_j=(c_j^{\dagger}+c_j)e^{\pm i\pi\sum_{l<j}c_l^{\dagger}c_l},\quad\sigma^x_j=2c_j^{\dagger}c_j-1.
\label{eq:JW}
\ee
with $\{c_i,c_j\}=\delta_{ij}$ and $\{c_i,c_j\}=\{c_i^{\dagger},c_j^{\dagger}\}=0$.
Consequently,
the disorder operator is mapped as $\mu_j=\prod_{k<j}\sigma^x_k=e^{\pm i\pi\sum_{k=1}^{j-1}c_k^{\dagger}c_k}$ and the relation $\sigma_j^z=(c_j^{\dagger}+c_j)\mu_j$ holds.

Furthermore,
the Majorana spinior $\psi = (\psi_L,\psi_R)^T$ is introduced with components \cite{KAROWSKI1978455}
\be
\psi_L(j)=\frac{(-1)^j}{\sqrt{a}}(c_j^{\dagger}e^{-i\pi/4}+c_je^{i\pi/4}),\quad
\psi_R(j)=\frac{(-1)^j}{\sqrt{a}}(c_j^{\dagger}e^{i\pi/4}+c_je^{-i\pi/4}),
\label{eq:Majorana}
\ee
where $a$ is the lattice spacing and
commutation relations follow
$\{\psi_R(i),\psi_R(j)\}=\{\psi_L(i),\psi_L(j)\}=\delta_{ij}/a$, $\{\psi_R(i),\psi_L(i)\}=0$.
Then the relations between spin operators and Majorana spinor are directly obtained as
\be
\sigma^x_j=-ai\psi_R(j)\psi_L(j),\quad
\sigma^z_j = (-1)^j\sqrt{\frac{a}{2}}\left(\psi_R(j)+\psi_L(j)\right)e^{\pm i\pi\sum_{l<j}(-ai\psi_R(l)\psi_L(l)+1)/2}.
\label{eq:sigmazpsi}
\ee
By taking the scaling limit $a\to0$ and $Ja$ being finite,
the critical TFIC can be decribed by the free Hamiltonian density in terms of Majorana spinor
\be
\mathcal{H}_{ms}=\psi^{\dagger}(x)\left(-i\gamma^5\frac{\partial}{\partial x}\psi(x)\right)
\ee
with $x=ja$, $\gamma^5=\sigma^z$ and $2Ja=1$.
In the field theory language,
$\sigma^z_j$, $\mu_j$ and $\sigma^x_j$ are referred to as order operator $\sigma(x)$, disorder operator $\mu(x)$ and energy operator $\epsilon(x)$, respectively.

Then we consider two copies of the critical transverse field Ising chains.
As the Majorana spinior is real,
two sets of Majorana spiniors $\psi^{(1,2)}$ obtained from the two chains are further gathered into complex Dirac spiniors $\chi$ as
\be
\chi = \frac{1}{\sqrt{2}}\left(\psi^{(1)}+i\psi^{(2)}\right),\quad
\chi^{\dagger} = \frac{1}{\sqrt{2}}\left(\psi^{(1)}-i\psi^{(2)}\right).
\ee
Then the Dirac spinors can be bosonized into free massless bosonic field following the bosonization rule
\be
\chi_R = \frac{\alpha_R}{\sqrt{N}}:e^{i\phi_R}:,\quad
\chi_L = \frac{\alpha_L}{\sqrt{N}}:e^{-i\phi_L}:,
\label{eq:rule}
\ee
where the normal ordering $::$ puts annihilation operators to the right and
$\alpha_{L,R}$ ensure the anti-commutation relation of $\chi_{R,L}$ \cite{2isingboson}.
Right- and left- going components of the bosonic field are introduced via the bisecting mode expansion of the free bosonic field
$\phi(x,t)=\phi_R(x-t)+\phi_L(x+t)$,
i.e.,
\be\ba
&\phi_R(t-x)=\frac{\phi_{0R}}{2\pi}-\frac{Q_R}{2N}(t-x)-\frac{i}{2\pi}\sum_{n\neq 0}\frac{\overline{a}_n}{n}e^{-2\pi in[(t-x)/N]}\\
&\phi_L(t+x)=\frac{\phi_{0L}}{2\pi}+\frac{Q_L}{2N}(t+x)-\frac{i}{2\pi}\sum_{n\neq 0}\frac{a_n}{n}e^{-2\pi in[(t+x)/N]},
\ea\ee
where $N$ is the system size,
and $Q_{R,L}$ are conjugation of the zero modes $\phi_{0R,L}$,
satisfying $[Q_R,\phi_{0R}]=-[Q_L,\phi_{0L}]=i/2$.
$a_{n}, \overline{a}_n$ are related to boson creation and annihilation operators ($\tilde{a}_n$ and $\tilde{a}_n^{\dagger}$) by
\be
a_n=\left\{
    \begin{aligned}
    -i\sqrt{n}\tilde{a}_n & (n>0)\\
    i\sqrt{-n}\tilde{a}_{-n}^{\dagger} & (n<0)\\
    \end{aligned}
    \right.
,\quad
\overline{a}_n=\left\{
    \begin{aligned}
    -i\sqrt{n}\tilde{a}_{-n} & (n>0)\\
    i\sqrt{-n}\tilde{a}_n^{\dagger} & (n<0)\\
    \end{aligned}
    \right.
,
   \ee
with non-vanishing commutators $[\overline{a}_n,\overline{a}_m]=[a_n,a_m]=n\delta_{n+m,0}$.
The dual field of $\phi$ is introduced as $\Theta(x,t)=\phi_R(x-t)-\phi_L(x+t)$
which satisfies
\be
\frac{\partial \Theta}{\partial x}=-\frac{\partial \phi}{\partial t}.
\label{eq:dual}
\ee
Thus we obtain the effective free bosonic field theory, namely $\mathcal{H}_{fb}=\partial^2 \phi(x)/\partial x^2$,
corresponding to a conformal field theory with central charge $c=1$ \cite{Ashkinteller}.

Taking the scaling limit of Eq.~\eqref{eq:sigmazpsi},
we have $\epsilon(x)^{(1,2)}=i\psi_R^{(1,2)}(x)\psi_L^{(1,2)}(x)$.
Several operator correspondences in the bosonization representation can be derived by
inserting the bosonization rules Eq.~\eqref{eq:rule} and using the Baker-Hausdorff formula for normal ordering operators \cite{Fradkin}
\be
:e^A::e^B:=e^{[A^+,B^-]}:e^{A+B}: \quad\text{if } [A^+,B^-]\text{ is c-number,}
\label{eq:BA}
\ee
where $\hat{O}=\hat{O}_++\hat{O}_-$ and $+/-$ denotes the creation/annihilation piece of the operator,
which are summarized in TABLE.~\ref{Tab:I} \cite{KADANOFF,2isingboson}.
\begin{table}[h]
    \setlength\tabcolsep{8pt}
    \renewcommand{\arraystretch}{1.25}
   \centering
    \begin{tabular}{ll}
    \specialrule{0em}{6pt}{6pt}
        \toprule [0.7pt]
        spin field & bosonized\\
        \midrule [0.4pt]
        $\sigma^{(1)}(x)\sigma^{(2)}(x)$ & $:\cos[\phi(x)/2]:$ \\
        $\epsilon^{(1)}(x)+\epsilon^{(2)}(x)$ & $:\cos[\phi(x)]:$\\
        $\epsilon^{(1)}(x)-\epsilon^{(2)}(x)$ & $:\cos[\Theta(x)]:$\\
        $\epsilon^{(1)}(x)\epsilon^{(2)}(x)$ & $\partial_\gamma\phi\partial^\gamma\phi \,(\gamma=x,\,t)$\\
        $\sigma^{(1)}(x)\mu^{(2)}(x)$ & $:\cos[\Theta(x)/2]:$\\
        $\mu^{(1)}(x)\mu^{(2)}(x)$ & $:\sin[\phi(x)/2]:$\\
        \bottomrule [0.7pt]
    \end{tabular}
    \caption{Bosonization correspondences for spin operators in the scaling limit.}
    \label{Tab:I}
\end{table}

Now we turn on weak interchain coupling,
the continuum theory of [Eq.(1a)] is given by the action
\be
\mathcal{A}_{\text{Ising}_h^2}=\mathcal{A}_{c=1/2}^{(1)}+\mathcal{A}_{c=1/2}^{(2)}+\lambda^{\prime}\int dxdt\sigma^{(1)}\sigma^{(2)},
\label{eq:isingh2action}
\ee
where each chain is described by a conformal field theory $\mathcal{A}_{c=1/2}^{(1,2)}$,
and the rescaled coupling strength $\lambda^{\prime}\propto\lambda^{7/4}$.
The last term in  Eq.~\eqref{eq:isingh2action} implies that additional interaction $\cos(\phi/2)$ should be added to the free boson theory [Eq.(3)],
which formally gives the Ising$_h^2$ action [Eq.(2)].
Eq.~\eqref{eq:isingh2action} appears in the same from as a sine-Gordon model,
while the difference lies in that the form is defined on $\mathbb{Z}_2$ orbifold \cite{coupleCFT}.

\section{B. The form factors}
\subsection{B.1 S-matrices and Form factors}
Distinct feature for integrable systems lies in the possiblity of deriving their exact scattering matrices,
moreover,
analytically obtaining the form factor series.
In this section,
we briefly introduce some basic concepts for integrble field theories following classic references \cite{10.1007/BFb0105279,Smirnov,Mussardobook}.

A single particle of mass $m$ is on-shell,
if satifying $E^2-q^2=m^2$ (the speed of light is set as 1),
where the energy and momentum can be parametrized in terms of rapidity $\theta$ as $E=m\cosh\theta$ and $q=m\sinh\theta$, respectively.
For forward particles,
$\theta$ runs on the real axis and by shifting to $i\pi+\theta$,
backward particle with the same rapidity is obtained.

Complete basis for a given integrable field theory includes vacuum $|0\rangle$ and $n$-particle asymptotic states ($n\in\mathbb{N}_+$),
the latter is denoting as
\be
|P_1(\theta_1)P_2(\theta_2)\cdots P_n(\theta_n)\rangle_{\text{in/out}},
\ee
with $j$-th particle of type $P$, travelling with rapidity $\theta_i$.
Such states are eigenstates of energy and momentum operators,
as $E=\sum_{j}E_j$ and $q=\sum_jq_j$.
In state, as denoted by the subscript,
is characterized as no interaction at infinite past,
meaning that the fastest particle is on the left and the slowest on the right,
which is opposite to the out state.
Thus we require $\theta_1>\theta_2>\cdots>\theta_n$ for in states and $\theta_1<\theta_2<\cdots<\theta_n$ for out states.

Existence of non-trivial set of local conserved charges,
as a basic feature of integrable systems,
highly restricted the particle scatterings:
no particle production through the scattering;
momenta are conserved between initial and final states;
any $n$ to $n$ scattering can be factorized into product of $2$ to $2$ scattering matrices (S-matrices).
Two-particle S-matrix can be written as
\be
|P_i(\theta_1)P_j(\theta_2)\rangle_{\text{in}}=S_{ij}^{kl}(\theta_1-\theta_2)|P_l(\theta_2)P_k(\theta_1)\rangle_{\text{out}},
\ee
where summation over $kl$ is implied and notice that S-matrix $S$ depends only on rapidity difference.
$S$ is a meromorphic function of $\theta$,
and physical stripe refers to $\text{Im} \theta\in[0,i\pi]$.
% Following relations should be satisfied:
% \be\ba
% &S(\theta)\text{ is real for } \theta \text{ purely imaginary};\\
% &\text{Unitarity: } S^{nm}_{ij}(\theta)S^{kl}_{nm}(-\theta) = \delta_i^k\delta_j^l;\\
% &\text{Crossing: } S^{kl}_{ij}(\theta)=S^{k\overline{j}}_{i\overline{l}}(i\pi-\theta).
% \ea\ee

$S(\theta)$ can have poles between $\theta=0$ and $\theta=i\pi$.
Bound states can be related to simple poles of S-matrices,
whose residue takes positive-real multiples of $i$ if considering forward channel.
\bigskip

Elementary form factors are defined for local operator $\hat{\mathcal{O}}$ as
\be
F_{\hat{\mathcal{O}}}^{P_1\cdots P_n}(\theta_1,\cdots,\theta_n)=\langle 0|\hat{\mathcal{O}}|P_1(\theta_1)\cdots P_n(\theta_n)\rangle_{\text{in}}.
\ee
In capacitance with scattering and conservation requirments,
following axioms holds:

1. Lorentz invariance
\be
F_{\hat{\mathcal{O}}}^{P_1\cdots P_n}(\theta_1+\lambda,\cdots,\theta_n+\lambda) =e^{s_{\hat{\mathcal{O}}}\lambda} F_{\hat{\mathcal{O}}}^{P_1\cdots P_n}(\theta_1,\cdots,\theta_n),
\ee
with Loreantz spin $s_{\hat{\mathcal{O}}}$ for $\hat{\mathcal{O}}$.

2. Exchange property
\be
F_{\hat{\mathcal{O}}}^{P_1\cdots P_n}(\theta_1,\cdots,\theta_j,\theta_{j+1},\cdots,\theta_n) = S(\theta_j-\theta_{j-1})F_{\hat{\mathcal{O}}}^{P_1\cdots P_n}(\theta_1,\cdots,\theta_{j+1},\theta_j,\cdots,\theta_n).
\ee

3. Cyclic property
\be
F_{\hat{\mathcal{O}}}^{P_1\cdots P_n}(\theta_1,\cdots,\theta_{n-1},\theta_n+2\pi i)=F_{\hat{\mathcal{O}}}^{P_1\cdots P_n}(\theta_n,\theta_1,\cdots,\theta_{n-1}).
\ee

4. Kinematical singularities
\be
-i\lim_{\tilde{\theta}\to\theta}(\tilde{\theta}-\theta)F_{\hat{\mathcal{O}}}^{P_1\cdots P_{n+2}}(\tilde{\theta},\theta,\theta_1,\cdots,\theta_n)=\left(1-\prod_{i=1}^nS(\theta-\theta_i)\right)F_{\hat{\mathcal{O}}}^{P_1\cdots P_n}(\theta_1,\cdots,\theta_n).
\ee

5. Bound state singularities
\be
-i\lim_{\theta_{ab}\to iU_{ab}^c}(\theta_{ab}-iU_{ab}^c)F_{\hat{\mathcal{O}}}^{P_1\cdots P_{n+2}}(\theta_a,\theta_b,\theta_1,\cdots,\theta_n)=\Gamma_{ab}^c F_{\hat{\mathcal{O}}}^{P_1\cdots P_{n+1}}(\theta_c,\theta_1,\cdots,\theta_n),
\ee
with $\theta_{ab}=\theta_a-\theta_b$, $\theta_c = \theta_a-i(\pi-U_{bc}^a)=\theta_b+i(\pi-U_{ac}^b)$ and $U_{ab}^c+U_{bc}^a+U_{ca}^b=2\pi$.
$iU_{ab}^c$ is the pole in $S(\theta)$ for the occurance of bound state $P_c$ in the scattering of $P_a$ and $P_b$,
with coupling strength $\Gamma_{ab}^c$ obtained by the expansion
\be
S(\theta\to iU_{ab}^c)\sim \frac{i(\Gamma_{ab}^c)^2}{\theta-iU_{ab}^c}.
\ee
Apart from the poles mentioned above,
$F_{\hat{\mathcal{O}}}$ is expected to be analytic for $0<\text{Im}\theta<2\pi$.

\subsection{B.2 Free field approach to the sine-Gordon form factor}
Due to the similarities,
before disscussing Ising$_h^2$ theory,
we review the form factors for the well-studied sine-Gordon model.
Here we follow the free field representation approach developed in \cite{Lukyanov1995}.
For convience,
we introduce the parameter $\xi=1/7$.
At this value, the sine-Gordon model contains soliton $A_{+1}$, antisoliton $A_{-1}$ and breathers $B_n$ with $n=1,2,\cdots<1/\xi$ and is reflectionless,
namely only diagonal scatterings would happen.
The lightest breather is formed as a bound state of soliton and anti-soliton pairs,
whilst heavier ones can be viewed as bound states of $B_1$s.

Following \cite{Jimbo1996},
the free boson $b(t)$ ($t\in\mathbb{R}$) is introduced,
whose commutation relation follows
\be
[b(t),b(t^{\prime})]=\frac{\sinh\frac{\pi t}{2}\sinh\pi t\sinh \frac{\pi t(\xi+1)}{2}}{t\sinh\frac{\pi t \xi}{2}}\delta(t+t^{\prime}).
\ee
The vacuum state $|\mbox{vac}\rangle$ for the Fock space is defined as $b(t)|\mbox{vac}\rangle=0$ for $t>0$.
We further introduce the vertex operators
\be\ba
&V(\theta)=e^{i\varphi(\theta)}=\mathcal{N}:e^{i\varphi(\theta)}:,\quad i\varphi(\theta)=\int_{-\infty}^{\infty}\frac{b(t)}{\sinh\pi t}e^{i\theta t}dt;\\
&\overline{V}(\theta)=e^{-i\overline{\varphi}(\theta)}=\overline{\mathcal{N}}:e^{-i\overline{\varphi}(\theta)}:,\quad i\overline{\varphi}(\theta)=\int_{-\infty}^{\infty}\frac{b(t)}{\sinh\frac{\pi t}{2}}e^{i\theta t}dt,
\ea \ee
with normalization constants $\mathcal{N}, \overline{\mathcal{N}}$.
From the integral representation we observe that $\overline{\varphi}(\theta)=\varphi(\theta+\frac{i \pi}{2})+\varphi(\theta-\frac{i \pi}{2})$.
Both $\varphi$ and $\overline{\varphi}$ are superpositions of creation and annihilation operators.

Using Eq.~\eqref{eq:BA}, we have
\be\ba
&V(\theta_1)V(\theta_2)=G(\theta_2-\theta_1):e^{i\varphi(\theta_1)+i\varphi(\theta_2)}:,\\
&V(\theta_1)\overline{V}(\theta_2)=W(\theta_2-\theta_1):e^{i\varphi(\theta_1)-i\overline{\varphi}(\theta_2)}:,\\
&\overline{V}(\theta_1)\overline{V}(\theta_2)=\overline{G}(\theta_2-\theta_1):e^{-i\overline{\varphi}(\theta_1)-i\overline{\varphi}(\theta_2)}:.
\label{eq:contraction}
\ea\ee
The explicit expressions for the exponentials of commutator $G, W$ and $\overline{G}$ will be discussed below.

The free field approach \cite{Lukyanov1995} leads to integral representations for
form factors of the vertex operators $e^{\pm i\phi}$ of the SG model.
Since breather form factors can be obtained from soliton-antisoliton form factors by the dynamical singularity axiom,
the asymptotic in-state with solitons and antisolitons provides a starting point \cite{Lukyanov1997}. 
Therefore, 
we consider
\be
\langle0|e^{il\phi}|A_{s_{1}}(\theta_1)\cdots A_{s_{2n}}(\theta_{2n})\rangle=\mathcal{E}_2\langle\langle Z_{s_{1}}(\theta_{1})\cdots Z_{s_{2n}}(\theta_{2n})\rangle\rangle,
\quad l=\pm 1,
\label{eq:ff element}
\ee
with vacuum expectation value $\mathcal{E}_2=\langle e^{il\phi}\rangle$ and the vertex operators are given by
\be\ba
Z_{+1}(\theta)&=\sqrt{\frac{iC_2}{4C_1}}e^{2l\theta}e^{i\varphi(\theta)},\\
Z_{-1}(\theta)&=\sqrt{\frac{iC_2}{4C_1}}e^{-2l\theta}\left\{e^{\frac{i\pi}{2\hat{\beta}^2}}\int_{C_+}\frac{d\gamma}{2\pi}e^{\left(-\frac{4l}{\hat{\beta}}-\frac{1}{\xi}\right )(\gamma-\theta)}e^{-i\overline{\varphi}(\gamma)}e^{i\varphi(\theta)}\right.\\
&\left.\qquad-e^{-\frac{i\pi}{2\hat{\beta}^2}}\int_{C_-}\frac{d\gamma}{2\pi}e^{\left(-\frac{4l}{\hat{\beta}}-\frac{1}{\xi}\right)(\gamma-\theta)}e^{i\varphi(\theta)}e^{-i\overline{\varphi}(\gamma)}\right\}.
\label{eq:ZF}
\ea \ee
Here $s_i\;(i = 1,\;\cdots\; 2n)$  is either $+1$ or $-1$, corresponding to soliton and antisoliton, respectively.
In addition,
$\sum_{i=1}^{2n}s_i=0$ as $e^{il\phi}$ preserves topological charge.
The integration contour $C_{+}$ ($C_-$) goes from $-\infty$ to $\infty$ with the pole $\gamma=\theta+i\pi/2$ ($\gamma=\theta-i\pi/2$) lying below (above) the contour.
The r.h.s. of Eq. \eqref{eq:ff element} contains vacuum expectation
of $2n$-product operators $\langle\langle \prod_{i}e^{\pm\hat{O}_i(\theta_i)}\rangle\rangle$ with $\hat{O}_i$ being either $\varphi$ or $\overline{\varphi}$ field.
Similar to Eq.~\eqref{eq:contraction},
vacuum expectation values can be calculated by Wick's theorem, \emph{e.g.},
\be\ba
&\langle\langle e^{i\varphi(\theta_1)}e^{-i\varphi(\theta_2)}e^{i\overline{\varphi}(\theta_3)}e^{-i\overline{\varphi}(\theta_4)}\rangle\rangle\\
=&\langle e^{-[i\varphi_+(\theta_1),i\varphi_-(\theta_2)]}:e^{i\varphi(\theta_1)}e^{-i\varphi(\theta_2)}::e^{i\overline{\varphi}(\theta_3)}::e^{-i\overline{\varphi}(\theta_4)}:\rangle\\
=&e^{-[i\varphi_+(\theta_1),i\varphi_-(\theta_2)]}e^{[i\varphi_+(\theta_1),i\overline{\varphi}_-(\theta_3)]} e^{-[i\varphi_+(\theta_1),i\overline{\varphi}_-(\theta_4)]}e^{-[i\varphi_+(\theta_2),i\overline{\varphi}_-(\theta_3)]}e^{[i\varphi_+(\theta_2),i\overline{\varphi}_-(\theta_4)]}e^{-[i\overline{\varphi}_+(\theta_3),i\overline{\varphi}_-(\theta_4)]}\\
=&\langle\langle e^{i\varphi(\theta_1)}e^{i\varphi(\theta_2)}\rangle\rangle^{-1}
\langle\langle e^{i\varphi(\theta_1)}e^{i\overline{\varphi}(\theta_3)}\rangle\rangle
\langle\langle e^{i\varphi(\theta_1)}e^{i\overline{\varphi}(\theta_4)}\rangle\rangle^{-1}
\langle\langle e^{i\varphi(\theta_2)}e^{i\overline{\varphi}(\theta_3)}\rangle\rangle^{-1}
\langle\langle e^{i\varphi(\theta_2)}e^{i\overline{\varphi}(\theta_4)}\rangle\rangle
\langle\langle e^{i\overline{\varphi}(\theta_3)}e^{i\overline{\varphi}(\theta_4)}\rangle\rangle.
\ea\ee

Explicit expressions for the required expectation values can be found in \cite{Lukyanov1995,Lukyanov1997},
which are summarized as follows.
\be\ba
&\langle\langle e^{i\varphi(\theta_2)}e^{i\varphi(\theta_1)}\rangle\rangle=G(\theta_1-\theta_2),\\
&\langle\langle e^{i\varphi(\theta_2)}e^{i\overline{\varphi}(\theta_1)}\rangle\rangle=\frac{1}{G(\theta_1-\theta_2-i\pi/2)G(\theta_1-\theta_2+i\pi/2)}=W(\theta_1-\theta_2),\\
&\langle\langle e^{i\overline{\varphi}(\theta_2)}e^{i\overline{\varphi}(\theta_1)}\rangle\rangle=\frac{1}{W(\theta_1-\theta_2-i\pi/2)W(\theta_1-\theta_2+i\pi/2)}=\overline{G}(\theta_1-\theta_2),\\
& G(\theta)=iC_1\sinh\left(\frac{\theta}{2}\right)\exp\left\{\int_0^{\infty}\frac{dt}{t}\frac{\sinh^2[t(1-i\theta/\pi)]\sinh[t(\xi-1)]}{\sinh(2t)\cosh(t)\sinh(t\xi)}\right\},\\
& W(\theta)=-\frac{2}{\cosh(\theta)}\exp\left\{-2\int_0^{\infty}\frac{dt}{t}\frac{\sinh^2[t(1-i\theta/\pi)]\sinh[t(\xi-1)]}{\sinh(2t)\sinh(t\xi)}\right\},\\
&\overline{G}(\theta)=-\frac{C_2}{4}\xi\sinh\left(\frac{\theta+i\pi}{\xi}\right)\sinh(\theta),\\
&C_1=\exp\left\{-\int_0^{\infty}\frac{dt}{t}\frac{\sinh^2(t/2)\sinh[t(\xi-1)]}{\sinh(2t)\cosh(t)\sinh(t\xi)}\right\}=G(-i\pi),\\
&C_2=\exp\left\{4\int_0^{\infty}\frac{dt}{t}\frac{\sinh^2(t/2)\sinh[t(\xi-1)]}{\sinh(2t)\sinh(t\xi)}\right\}=\frac{4}{[W(i\pi/2)\xi\sin(\pi/\xi)]^2}.
\label{eq:GWC}
\ea\ee
It is worth noting that the integral formulae for $G(\theta)$ and $W(\theta)$ are convergent for $-2\pi-\pi\xi<\mbox{Im}[\theta]<\pi\xi$ and $-2\pi+2\pi\xi<\mbox{Im}[\theta]<-2\pi\xi$, respectively.
Their analytic continuations will be discussed later.

\subsection{B.3 Form factors of $\cos\phi$ in Ising\(_h^2\) theory}
\subsubsection{Soliton-antisoliton in-state}
The SG$_{1/2}$ form factor with soliton-antisoliton in-state reads \cite{Lukyanov1997}
\be \ba
F_{A_{s}A_{-s}}^{\exp(i\phi)}(\theta_2,\theta_1)|_{\text{SG}_{1/2}}
&=\langle0|e^{i\phi}|A_{s}(\theta_2)A_{-s}(\theta_1)\rangle_{\text{SG}_{1/2}}
=\mathcal{E}_2\frac{G(\theta_2-\theta_1)}{G(-i\pi)}\frac{4i e^{s\frac{\theta_2-\theta_1+i\pi}{2\xi}}}{\xi\sinh\left(\frac{\theta_2-\theta_1+i\pi}{\xi}\right)}\\
&\times\left[\cosh\frac{(\theta_2-\theta_1)}{2}\cot\frac{\pi\xi}{2}\cot\frac{\pi\xi}{2}\cot\pi\xi
+\cosh\frac{3(\theta_2-\theta_1)}{2}\cot\frac{\pi\xi}{2}\cot\pi\xi\cot\frac{3\pi\xi}{2}\right],
\label{eq:phiff_sas}
\ea \ee
and $F_{A_{s}A_{-s}}^{\exp(i\phi)}(\theta_2,\theta_1)|_{\text{SG}_{1/2}}=F_{A_{-s}A_{s}}^{\exp(-i\phi)}(\theta_2,\theta_1)|_{\text{SG}_{1/2}}$.
The S-matrix elements of SG$_{1/2}$ and Ising$_h^2$ theories are related by $S_{A_{+1}A_{-1}}|_{\text{SG}_{1/2}}=-S_{A_{+1}A_{-1}}|_{\text{Ising}_h^2}$.
Assuming that the two form factors are related by $F_{A_{-s}A_{s}}^{\cos\phi}(\theta_1,\theta_2)|_{\text{Ising}_h^2}=f(\theta_1-\theta_2)F_{A_{-s}A_{s}}^{\cos\phi}(\theta_1-\theta_2)|_{\text{SG}_{1/2}}$, form factor axioms \cite{Mussardobook} indicate that $f(\theta)$ should satisfy
\be
f(\theta)=-f(-\theta),\quad
f(i\pi-\theta)=f(i\pi+\theta).
\ee
It is obvious that the minimal solution of the above equation is simply $f(\theta)=\sinh\left(\theta/2\right)$. Since we do not expect any new singularities from physical considerations, we take this minimal solution.

\subsubsection{Breather(s) in-state}
As breather S-matirces of SG$_{1/2}$ and Ising$_h^2$ are identical,
the SG$_{1/2}$ form factors can be directly applied for the latter case.
Form factors for in-state containing breather(s) can be derived from dynamical pole of soliton-antisoliton case.
On the other side,
S-matrix of $B_1B_1$ coincides with that of the Zamolochikov-Faddeev algebra generated by a current operator in the deformed Virasoro algebra \cite{virasoro},
which provides a simple formalism to calculate form factors for breather(s) in-state.
The bound state pole gives \cite{Jimbo1996,Lukyanov1997}
\be\ba
Y(\theta^{\prime}=\theta_1-i\pi(1-\xi)/2) &= \mbox{Res}_{\theta_2=\theta_1-i\pi(1-\xi)} Z_{+1}(\theta_2)Z_{-1}(\theta_1)\\
&=\frac{i\Gamma_{-+}^1\lambda}{2\sin(\pi\xi)}\left\{e^{-2i\pi\xi}e^{-iw(\theta_1+i\pi(1-\xi)/2+i\pi/2)}-e^{2i\pi\xi}e^{iw(\theta_1+i\pi(1-\xi)/2-i\pi/2)}\right\},
\ea\ee
with coupling strength $\Gamma_{+-}^1=\sqrt{2\cot(\pi/14)}$ and
\be \ba
&w(\theta)=\varphi(\theta+i\pi\xi/2)-\varphi(\theta-i\pi\xi/2),\\
&\lambda=2\cos(\pi\xi/2)\sqrt{2\sin(\pi\xi/2)}\exp\left\{-\int_0^{\pi\xi}dt\frac{t}{2\pi \sin t}\right\}.
\ea\ee
The expectation value $\langle\langle e^{iw(\theta_1)}e^{iw(\theta_2)}\rangle\rangle$ can be calculated from the formulae in the previous section.
Taking $\langle\langle e^{i\omega(\theta)}\rangle\rangle=1$, we have
\be\ba
R(\theta)=\langle\langle e^{iw(\theta_1)}e^{iw(\theta_2)}\rangle\rangle
&=C_3 \exp\left\{8\int_0^{\infty}\frac{dt}{t}\frac{\sinh(t)\sinh(t\xi)\sinh[t(1+\xi)]}{\sinh^2(2t)}\sinh^2\left[t\left(1-\frac{i\theta}{\pi}\right)\right]\right\},\\
& C_3=\exp\left\{4\int_0^{\infty}\frac{dt}{t}\frac{\sinh(t)\sinh(t\xi)\sinh[t(1+\xi)]}{\sinh^2(2t)}\right\},
\label{eq:R}
\ea\ee
for $\mbox{Im}[\theta]\in(-2\pi+\pi\xi,-\pi\xi)$.
The general formula for $nB_1$ form factor of $e^{is\phi}$ ($s=\pm 1$) reads \cite{Takacs2010}
\be\ba
\langle0|e^{is\phi}|B_1(\theta_1)\cdots B_1(\theta_n)\rangle
=\mathcal{E}_2 [2s]_{\xi}(i\lambda)^n\prod_{i<j}\frac{R(\theta_j-\theta_i)}{e^{\theta_i}+e^{\theta^j}}Q^{(n)}(e^{\theta_1},\cdots,e^{\theta_n}),
\ea\ee
with
\be\ba
&[z]_{\xi}=\frac{\sin\pi\xi z}{\sin\pi\xi},\qquad
Q^{(1)}=1,\\
& Q^{(n)}(e^{\theta_1},\cdots,e^{\theta_n})=\det[2s+i-j]_{\xi}\sigma^{(n)}_{\xi}(e^{\theta_1},\cdots,e^{\theta_n})_{i,j=1,\cdots,n-1}\quad\mbox{for}\,\,n>1,
\ea\ee
and $\sigma^{(n)}_{m}$ denotes the symmetric polynomial defined by
\be
\prod_{i=1}^n(z+z_i)=\sum_{m=0}^n z^{n-m}\sigma_{m}^{(n)}(z_1,\cdots,z_n).
\ee
For example,
\be
\langle 0|e^{i\phi}|B_1(\theta_1)B_1(\theta_2)\rangle=\mathcal{E}_2\left(\frac{\sin2\pi\xi}{\sin\pi\xi}\right)^2(i\lambda)^2R(\theta_2-\theta_1).
\ee

Form factor contains heavier breathers can be derived from the bootstrap procedure \cite{Mussardo},
as $B_{n}$ is a bound state of $B_{n-1}$ and $B_1$.
As a result, we have
\be\ba
&\langle0|e^{is\phi}|B_n(\theta_n)B_k(\theta_k)\cdots B_l(\theta_l)\rangle\\
=&\Gamma_{1,1}^2\Gamma_{1,2}^3\cdots\Gamma_{1,n-1}^n\langle0|e^{is\phi}|\underbrace{B_1(\theta_n+\frac{1-n}{2}i\pi\xi)B_1(\theta_n+\frac{3-n}{2}i\pi\xi)\cdots B_1(\theta_n+\frac{n-1}{2}i\pi\xi)}_{n(\geq 2)\,B_1} B_{k}(\theta_k)\cdots B_l(\theta_l)\rangle,
\label{eq:ffphi_B}
\ea\ee
where the coupling strength for $B_1B_{n-1}\to B_n$ is
\be
\Gamma_{1,n-1}^n=\sqrt{\frac{2\tan\frac{(n-1)\pi\xi}{2}\tan\frac{n\pi\xi}{2}}{\tan\frac{\pi\xi}{2}}}.
\ee
As a consistent check for the selection rule of $\cos\phi$,
notice that $[2s]_{\xi}=-[-2s]_{\xi}$ and $\det[2s+i-j]_{\xi}=(-1)^n\det[2s+i-j]_{\xi}$ for $i,j=1,\dots,n-1$,
which leads to vanishing $\langle 0|\cos\phi|B_{1,3,5}\rangle$.
Then selection rule for multi-breathers in-state follows.

\subsection{B.4 Form factors of \(\cos\Theta\) in Ising\(_h^2\) theory}
$e^{\pm i\Theta}$ is identified with charge-1 raising/lowering operator in \cite{chargeraising} using the relation Eq.~\eqref{eq:dual}.
The $\cos\Theta$ form factors can be derived following the same process as Eq.~\eqref{eq:ff element}, with a different normalization constant.
For example,
in the presence of one antisoliton,
\be\ba
&\langle0|e^{i\Theta}|A_{+1}(\theta_3)A_{-1}(\theta_2)A_{+1}(\theta_{1})\rangle\\
=&\frac{iC_2\sqrt{\mathcal{Z}_1(0)}}{4C_1}\prod_{i<j}G(\theta_i-\theta_j)\left\{e^{\frac{i\pi^2}{2\beta^2}}\int_{C_+}\frac{d\gamma}{2\pi}e^{(\theta_k-\gamma)\frac{1}{\xi}}\prod_{l=1}^k W(\gamma-\theta_l)\prod_{l=k+1}^{3}W(\gamma-\theta_l)\right.\\
&-\left.e^{-\frac{i\pi}{2\beta^2}}\int_{C_-}\frac{d\gamma}{2\pi}e^{(\theta_k-\gamma)\frac{1}{\xi}}\prod_{l=1}^{k-1}W(\theta_l-\gamma)\prod_{l=k}^{3}W(\gamma-\theta_l)\right\}\,.
\ea\ee
The explicit expression for the normalization operator $\sqrt{\mathcal{Z}_{\pm 1}(0)}$ is given in \cite{chargeraising} and the contour $C_{\pm}$ follow the same convention as in Section I.

In the energy region and particle channels of interest, we will focus on single (anti)soliton and and (anti)soliton---$B_n$ states.
Since no $A_{\pm 1}A_{\pm 1\;{\text or}\; \mp 1}$ scatterings will be encountered in the in-state,
again,
the SG$_{1/2}$ form factor applies for that of Ising$_h^2$ theory.
Furthermore,
by applying charge conjugation transformation $\mathcal{C}$ we observe that
\be
\langle0|e^{i\Theta}|A_{+}(\theta_2)B_{m}(\theta_1)\rangle
=\langle0|\mathcal{C}\mathcal{C}^{\dagger}e^{i\Theta}\mathcal{C}\mathcal{C}^{\dagger}|A_{+}(\theta_2)B_{m}(\theta_1)\rangle
=\langle0|e^{-i\Theta}|A_{-1}(\theta_2)B_{m}(\theta_1)\rangle.
\ee
Thus we only consider positive charged states,
without encountering the complex integral form for $Z_-$ [Eq.~\eqref{eq:ZF}].

For single soliton in-state,
\be
\langle0|e^{i\Theta}|A_{+1}(\theta_1)\rangle=\sqrt{\mathcal{Z}_{1}(0)}.
\ee

And for soliton-$B_n$ in-state,
\be \ba
&\langle 0|e^{i\Theta}|A_{+1}(\theta_2)B_{n}(\theta_1)\rangle\\
=&\Gamma^1_{1,1}\Gamma^3_{1,2}\cdots\Gamma^n_{1,n-1}\langle 0|e^{i\Theta}|A_{+1}(\theta_2)\underbrace{B_1(\theta_1+\frac{1-n}{2}i\pi\xi)B_1(\theta_1+\frac{3-n}{2}i\pi\xi)\cdots B_1(\theta_1+\frac{n-1}{2}i\pi\xi)}_{n\,B_1} \rangle\\
&=\Gamma^1_{1,1}\Gamma^3_{1,2}\cdots\Gamma^n_{1,n-1}\sqrt{\mathcal{Z}_{1}(0)}\langle\langle Z_{+1}(\theta_2)Y(\theta_1+\frac{1-n}{2}\pi\xi)Y(\theta_1+\frac{3-n}{2}\pi\xi)\cdots Y(\theta_1+\frac{n-1}{2}\pi\xi)\rangle\rangle.
\ea \ee
Wick's theorem can be applied to compute form factors of $e^{\pm i\varphi}$, $e^{\pm i\omega}$ operators,
leading to the final results.

\subsection{B.5 Analytic continuations}
As mentioned before,
the integral expressions for $G(\theta)$, $W(\theta)$ and $R(\theta)$ are convergent for some $\mbox{Im}[\theta]$ range.
Here we collected the analytic continuation relations between different ranges of $\mbox{Im}[\theta]$ for these functions,
and also their alternative expressions.

The integral formula Eq.~\eqref{eq:R} for $R(\theta)$ implies \cite{Lukyanov1997}
\be
R(\theta)R(\theta\pm i\pi)=\frac{\sinh(\theta)}{\sinh(\theta)\mp i\sin(\pi\xi)},
\ee
which can be applied to calculate $R(\theta)$ from non-convergent $\mbox{Im}[\theta]$ by shifting $i\pi$ recursively.
Also,
the continuation relation for $W(\theta)$ is derived as \cite{Lukyanov1995}
\be \ba
W(\theta-i\pi)&=W(-\theta-i\pi)\\
\frac{W(\theta-\frac{i\pi}{2})W(\theta+\frac{i\pi}{2})}{W^2(\frac{i\pi}{2})}&=-\frac{\xi \sin^2(\frac{\pi}{\xi})}{\sinh\theta\sinh\frac{\theta+i\pi}{2}}.
\label{eq:ana W}
\ea\ee
From Eq. \eqref{eq:ana W}, the continuation relation for $G(\theta)$ is direct, which reads as
\be
\frac{1}{G(\theta-i\pi)G(\theta+i\pi)G(\theta)^2}=-\frac{4}{C_2\xi\sinh\theta\sinh\frac{\theta+i\pi}{2}},
\ee
where the second expression for $C_2$ in Eq.~\eqref{eq:GWC} is inserted.

For convenience,
here we list some alternative expressions for these function from \cite{Takacs2010,Takacs2011,Palmai2012},
which deals the integral with additional exponential factor and has no restriction on $\theta$.
\be\ba
R(\theta)=v(i\pi+\theta,-1)v(i\pi&+\theta,-\xi)v(i\pi+\theta,1+\xi)v(-i\pi-\theta,-1)v(-i\pi-\theta,-\xi)v(-i\pi-\theta,1+\xi)\\
v(\theta,\zeta)&=\prod_{k=1}^N \left(\frac{\theta+i\pi(2k+\zeta)}{\theta+i\pi(2k-\zeta)}\right)^k\exp\left\{\int_0^{\infty}\frac{dt}{t}\left(-\frac{\zeta}{4\sinh\frac{t}{2}}-\frac{i\zeta\theta}{2\pi\cosh\frac{t}{2}}\right.\right.\\
&\left.\left.+(N+1-Ne^{-2t})e^{-2Nt+\frac{it\theta}{\pi}}\frac{\sinh\zeta t}{2\sinh^2t}\right)\right\}.
\ea\ee
\be\ba
W(\theta)&=-\frac{2}{\cosh\theta}\prod_{k=1}^N\frac{\Gamma\left(1+\frac{2k-\frac{5}{2}+\frac{i\theta}{\pi}}{\xi}\right)\Gamma\left(1+\frac{2k-\frac{5}{2}-\frac{i\theta}{\pi}}{\xi}\right)\Gamma\left(\frac{2k-\frac{1}{2}}{\xi}\right)^2}{\Gamma\left(1+\frac{2k-\frac{3}{2}}{\xi}\right)^2\Gamma\left(1+\frac{2k+\frac{1}{2}-\frac{i\theta}{\pi}}{\xi}\right)\Gamma\left(1+\frac{2k-\frac{3}{2}+\frac{i\theta}{\pi}}{\xi}\right)}\\
&\times\exp\left\{-2\int_0^{\infty}\frac{dt}{t}\frac{e^{-4Nt}\sinh^2\left[t\left(1-\frac{i\theta}{\pi}\right)\right]\sinh[t(\xi-1)]}{\sinh2t\sinh\xi t}\right\}
\ea\ee
\be\ba
G(\theta)=iC_1\sinh\left(\frac{\theta}{2}\right)\prod_{k=1}^N \tilde{g}(\theta,\xi,k)^k&\exp\left\{\frac{dt}{t}e^{-4Nt}(1+N-Ne^{-4t})\sinh^2\left[t\left(1-\frac{i\theta}{\pi}\right)\right]\frac{\sinh[t(\xi-1)]}{\sinh(2t)\cosh(t)\sinh(t\xi)}\right\}\\
\tilde{g}(\theta,\xi,k)&=\frac{\Gamma\left(\frac{(2k+1+\xi)\pi-i\theta}{\pi\xi}\right)\Gamma\left(\frac{2k+1}{\xi}\right)^2\Gamma\left(\frac{(2k+1)\pi-i\theta}{\pi\xi}\right)}{\Gamma\left(\frac{2k+\xi}{\xi}\right)^2\Gamma\left(\frac{(2k+\xi)\pi-i\theta}{\pi\xi}\right)\Gamma\left(\frac{(2k-2+\xi)\pi+i\theta}{\pi\xi}\right)}.
\ea\ee
All expressions above are independent of the choice of integer $N$.

\section{C. The $S^{y}$ and $S^{z}$ channels}
The $S^{y}$ and $S^{z}$ channels are not independent.
By extending the proof in Ref.~\onlinecite{JD2014}, it can be established that
\be
4h^2 D_{S_j^y}(\omega, q)=\omega^2D_{S_j^z}(\omega, q)
\label{eq:zzandyy}
\ee
for the quantum Ising ladder, 
where $h$ denotes the strength of external field coupling on $S_j^x$,
Eq.~(\ref{eq:zzandyy}) is generally valid
for any $d$-dimensional Ising model with anisotropic spin coupling
strength and arbitrary interaction range in the
presence of both transverse and longitudinal fields.
It also implies that $D_{S_j^z}(\omega)$ must converge faster than $\omega^{-3}$ as $\omega\to\infty$
due to sum rule constraint.
The ratio $D_{S_j^y}/D_{S_j^z}$ predicted from Eq.~\eqref{eq:zzandyy} aligns with that shown in Fig.~4,
which is suppressed for low energy while enhanced for high energy.
The crossing point in Fig.~4 appears at $\omega = 2h =  2g J$.

\begin{figure}[h!]
    \centering
    \includegraphics[width=0.35\textwidth]{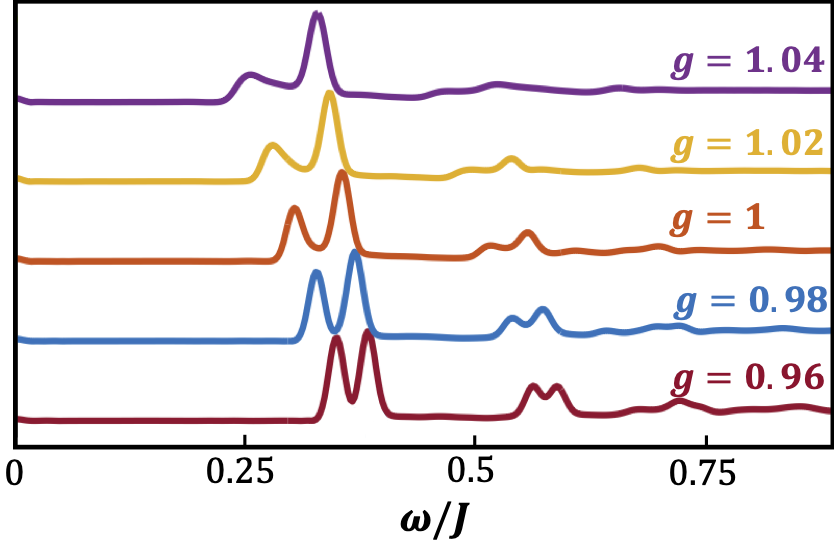}
    \caption{$D_{S^{(1,2)x}}(q=0)$ spectra from iTEBD calculation for the transverse field Ising
    ladder with $\lambda=0.1J$ and transverse fields at fine-tuning point and neighboring values.
    }
    \label{fig:gneq1}
\end{figure}

\section{D. Spectra deviating from integrable point}
\begin{figure}[h!]
    \centering
    \includegraphics[width=0.8\textwidth]{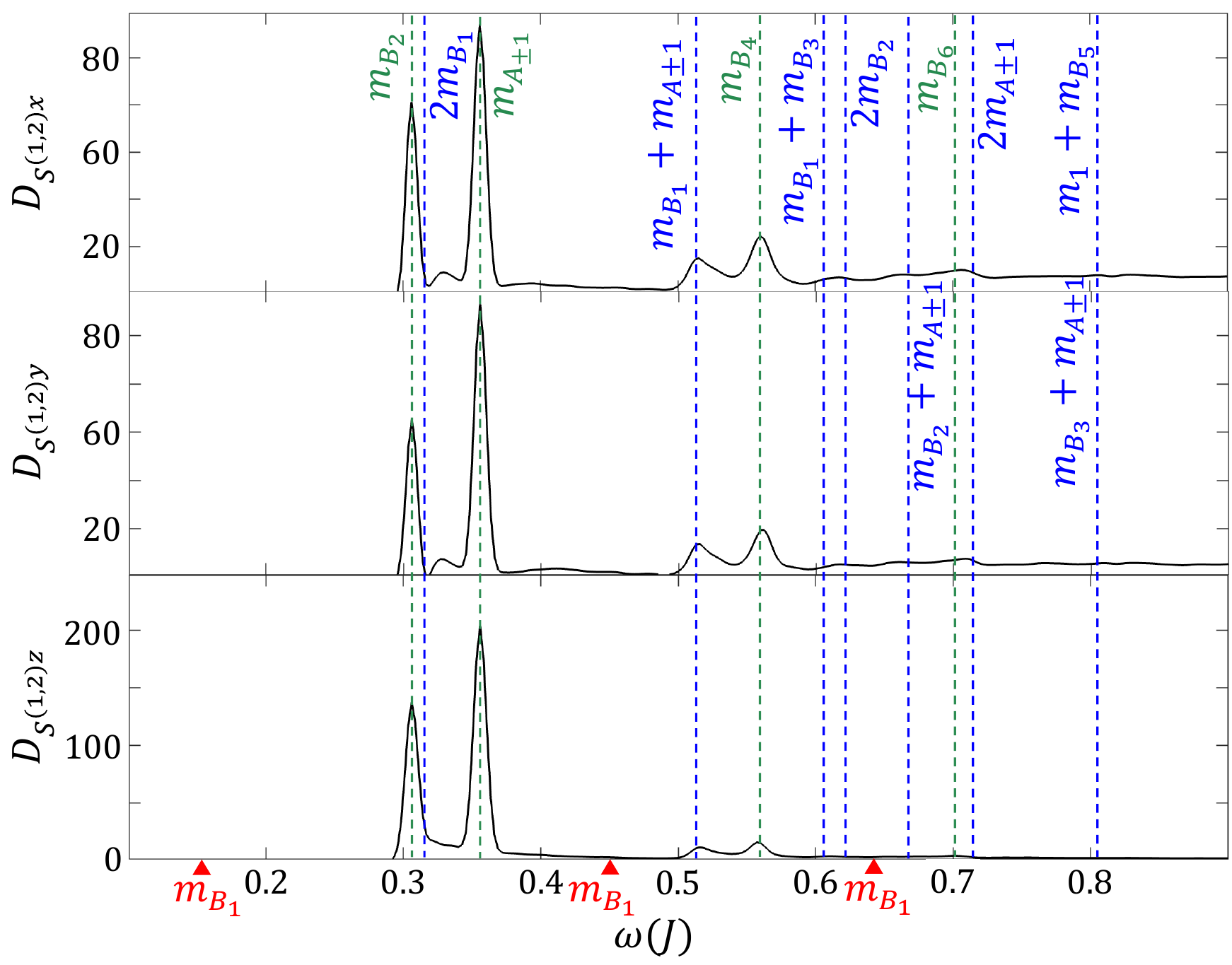}
    \caption{$D_{S^{x,y,z}}(q=0)$ spectra from iTEBD calculation for the transverse field Ising
    ladder with perturbative longitudinal field $h_z=0.005J$. Green dashed lines mark
    single-$D_8^{1}$-particle resonant peaks, while blue dashed lines mark the
    two-$D_8^{1}$-particle thresholds. The perturbation field does not
    alternate the $D_8^{1}$ physics in the quantum Ising$_h^2$ integrable field theory.
    }
    \label{fig:hz}
\end{figure}

The Ising$_h^2$ physics governs the neighboring region of the integrable point.
When the transverse field is slightly tuned away from $g_0=1$,
the model immediately becomes non-integrable,
which appears as
\be
H_{\text{pert}} = H_{\text{Ising}_h^2}+ \lambda\int dx~ \cos\phi(x),
\ee
with $\lambda$ being the rescaled perturbation strength $g-g_0$,
which has the same form as a double-frequency sine-Gordon theory \cite{DELFINO1998,PhysRevB.102.014426}.
Field theory perturbation \cite{marton2019} can be carried out in this region,
leading to perturbed ground state, single particles and so on (denote by $\sim$).
For examples,
\be\ba
|\tilde{0}\rangle&=|0\rangle-\lambda\left\{\sum_{j=2,4,6}\frac{\langle B_j(0)|\cos\phi|0\rangle}{m_{B_j}}|B_j(0)\rangle+\int\frac{d\theta}{2\pi}\frac{\langle B_1(\theta)B_1(-\theta)|\cos\phi|0\rangle}{2m_{B_1}\cosh\theta}|B_1(\theta)B_1(-\theta)\rangle+\cdots\right\};\\
|\tilde{B}_1(0)\rangle&=|B_1(0)\rangle-\lambda\left\{\sum_{j=3,5}\frac{\langle B_j(0)|\cos\phi|B_1(0)\rangle}{m_{B_j}-m_{B_1}}|B_j(0)\rangle\right.\\&+\left.\sum_{s=\pm1}\int\frac{d\theta}{2\pi}\frac{\langle A_s(\theta)A_{-s}(-\theta)|\cos\phi|B_1(0)\rangle}{2m_{A_{\pm1}}\cosh\theta}|A_s(\theta)A_{-s}(-\theta)\rangle+\cdots\right\}.
\ea\ee
The dark property of perturbed $B_1$ can be viewed from the DSF channel $\langle\tilde{0}|\cos\phi|\tilde{B}_1\rangle$.
Parity requirement forbids most of the transitions between the perturbation components,
and the contributions from $|A_{+1}A_{-1}\rangle$, $|A_{-1}A_{+1}\rangle$ in $\tilde{B}_1$,
which always appear in pairs,
cancel with each other as can be observed by applying charge parity conjugation $\mathcal{C}$.
It implies the absent DSF of $\tilde{B}_1$ from the perturbed ground state $|\tilde{0}\rangle$,
exhibiting the preservation of the dark property of $\tilde{B}_1$,
which is also numerically confirmed with different transverse field strength [Fig.~\ref{fig:gneq1}].
Furthermore,
perturbation that couples to Ising spin as $J\sum_{j}h_z\sigma^{z(1,2)}_j$ results in
the same Ising$_h^2$ dominant physics [Fig.~\ref{fig:hz}].

\section{E. The iTEBD simulations}
In this section, we provide illustrations on the numerical method used in calculating DSF in the main text.
We calculate the zero-temperature space-time correlation by employing the iTEBD method[ref],
which can be generally expressed as 
\be \ba
\langle\mathcal{O}(x,t)\mathcal{O}(0,0)\rangle = \langle\psi_{GS}|e^{-itH}\mathcal{O}(x,0)e^{itH}\mathcal{O}(0,0)|\psi_{GS}\rangle =\langle\psi_{GS}|e^{-itH}\mathcal{O}(x,0)e^{itH}|\tilde{\psi}\rangle,
\label{eq:realtime}
\ea \ee
where $|\psi_{GS}\rangle$ is the ground state, $\mathcal{O}(0,0)|\psi_{GS}\rangle=|\tilde{\psi}\rangle$ and $\mathcal{O}$ refers to a local observable.
Standard real-time evolution can be applied to the matrix product states $|\psi\rangle$ and $|\tilde{\psi}\rangle$.
Then the DSF can be obtained by performing the Fourier transformation on Eq.~\eqref{eq:realtime}.

In practical calculations, we employ the infinite MPS with the assumption of translational symmetry in the ground state, and control the entanglement entropy (computational accuracy) of systems by adjusting the truncation dimension $D$. 
A sixth-order Suzuki-Trotter decomposition is applied to minimize time-step errors \cite{Hatano2005}.
All the systems involved in this work are gapped Ising systems so that a truncation dimension of D=40 is sufficient to ensure the accuracy of the ground state calculations (with a truncation error less than $10^{-10}$). For the calculation of  space-time correlations, the truncation error gradually increases with evolution time.
To balance computing resources and acceptable errors, we choose a truncation dimension $D=160$  for the computation, ensuring the truncation error less than $10^{-4}$.
We set the time-step $\tau = 0.6/J$ and the number of steps $M=1200$.

\begin{figure}[htp]
    \centering
    \includegraphics[width=0.75\textwidth]{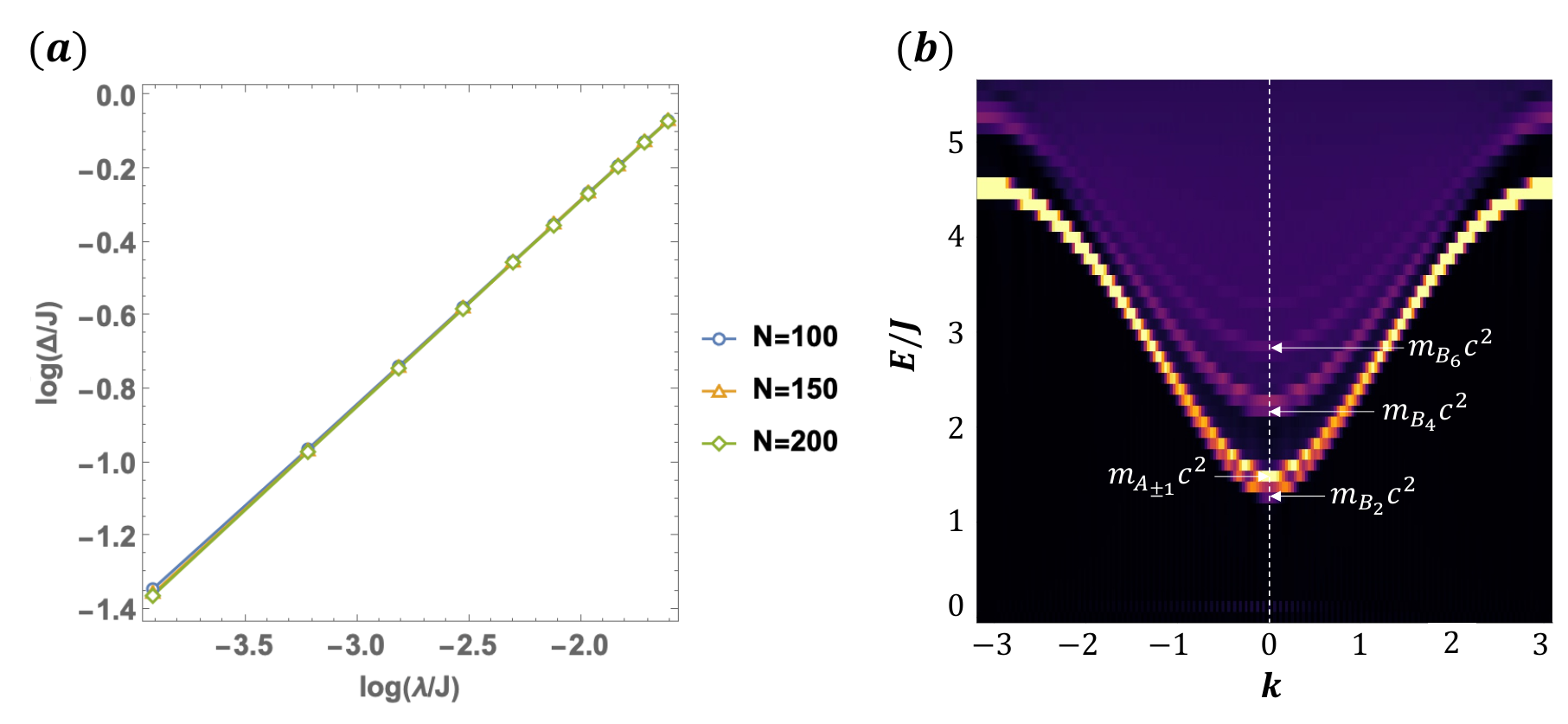}
    \caption{(a) Relation between $\log\lambda$ and $\log\Delta$ for different system sizes calculated by DMRG \cite{ITensor,ITensor-r0.3},
    with truncation error cutoff setting as $10^{-12}$ and maxmium bond dimension $400$.
    By fitting $\log\Delta\propto\zeta\log\lambda$, $\zeta$ is obtained as $0.555414$, $0.560421$ and $0.562057$ for
    $N=100$, $150$ and $200$, respectively.
    (b) $S^x$ DSF for Hamiltonian Eq.~(1a) calculated with $N=200$ and $\lambda=0.1J$.}
    \label{fig:scaling}
\end{figure}
\section{F. Numerical simulations on the finite-size system}
The Ising$_h^2$ integrable field theory emerges in the continous limit.
However,
characteristic features and the dark particles can be realized in discrete systems with intermediate system sizes,
facilitating direct simulations in cold atom systems and etc.

Following the field theory,
we have the scaling relation for the gap $\Delta\sim \lambda^{4/7}$.
Fig.~\ref{fig:scaling}a shows the density matrix renormalization group (DMRG) result for $\Delta$ obtained
in finite size systems with Hamiltonian [Eig.~(1a)].
The scaling relation is comparable with the field theory prediction.
DSF for the system of length $200$ is shown in Fig.~\ref{fig:scaling}b. The obtained energy for the first band at zone center ($1.3J$, identified as $m_{B_2}c^2$) through local spin excitation is about 2 times of the gap $\Delta=0.6353J$ (identified as $m_{B_1}c^2$), 
which is consistent with the IIFT result.

\bibliography{Isingh2}

\end{document}